\documentclass{aastex701}

\usepackage{float}
\usepackage{subcaption}
\usepackage{booktabs}
\usepackage{tabularx}
\usepackage{amsmath,amssymb}
\usepackage{array}
\usepackage{upgreek}

\begin{document}

\title{Retrieving Ocean Glint Reflectance Signatures from Directly Imaged Earth-like Exoplanets}

\author[orcid=0000-0000-0000-0001,sname='Cornish']{Eleanor Cornish}
\affiliation{University of Arizona, Department of Astronomy}
\email[show]{eleanorcornish@arizona.edu}  

\author[orcid=0000-0000-0000-0002,gname=Tyler, sname='Robinson'] {Tyler Robinson}
\affiliation{University of Arizona, Lunar and Planetary Laboratory}
\email[show]{tdrobin@arizona.edu}

\begin{abstract}
Future space telescopes searching for extraterrestrial life require methods to detect environments that can potentially support life. Liquid water is considered essential for life as we know it and its presence is therefore an important observable when characterizing potentially Earth-like exoplanets. Remotely inferring the presence of surface liquid water on a world from spatially unresolved observations is an outstanding challenge where the detection of ocean glint could provide supporting evidence. This study assesses the feasibility of detecting ocean glint from Earth-like exoplanets using simulated data from NASA's Habitable World Observatory (HWO). We conducted retrievals on these simulated observations using a model that included a glint contribution and performed model selection against retrievals that did not include glint. We find that over a range of spectral signal-to-noise ratios from 5 to 25, a phase angle of at least $120^{\circ}$ is necessary for glint detection. Our results offer the opportunity to shape mission development by providing constraints on key telescope observing requirements. 
\end{abstract}



\section{Introduction}

Identifying the presence of surface liquid water on an exoplanet would have major implications for our ongoing search for life beyond Earth. One technology that could allow for the identification of oceans on exoplanets is reflected light direct imaging, which is a planned capability for NASA's Habitable Worlds Observatory (HWO). Most generally, HWO will use high contrast imaging techniques to study targets in reflected light with enough sensitivity to observe unresolved Earth-like exoplanets and characterize these worlds for signs of habitability and life \citep{feinberg2024habitable}. The HWO mission will likely be capable of direct imaging observations spanning visible to near-infrared wavelengths.

Ocean glint\,---\,the specular reflection of sunlight from surface liquid water\,---\,is an observable that a direct imaging mission could potentially use to assess planetary habitability. A characteristic of glint is that liquid surfaces exhibit relatively weak specular reflection at near-normal illumination angles and increasingly strong specular reflection at grazing illumination angles. In the context of exoplanet observations, this corresponds to an observable brightening effect at large phase angles (planet-star-observer angle) due to increased reflectance at glancing angles \citep{mccullough2006modelspolarizedlightoceans,2008Icar..195..927W}. Consequently, the glint effect is most pronounced when an Earth-like world is observed at crescent phase, where models have predicted that ocean glint can more than double Earth's brightness at some wavelengths \citep{robinson2010detecting,vaughan2023chasing}. Thus, glint effects are weak at full phase through quadrature (i.e., a phase angle of 90 degrees, or half-illumination), and grow with phase angle through crescent phases. 

Glint has previously been used to identify surface liquids on other celestial bodies. Titan's methane oceans were identified using glint detection methods \citep{stephan2010specular}, and further studies have used these glints to characterize Titan's seas \citep{barnes2014cassini}. Future exoplanet observations will differ from Titan glint observations, as Titan being about 10 times further from the Sun than Earth implies that its glint features are about an order of magnitude smaller (in diameter) than Earth's. Thus, even at wavelengths with surface sensitivity, Titan sea glints only contribute a small amount of flux to disk-averaged observations of the moon. Despite these differences, Titan still acts as a promising case study that supports the idea that glint signatures can indicate the presence of surface liquids.

When considering glint detectability, it is important to remember that the range of accessible phase angles for a target depends on its apparent orbital inclination. Only certain orbital geometries produce observable glint \citep{vaughan2023chasing}. For example for a face-on orbit, $i=0^{\circ}$, we would not observe orbital phase variations, and thus could not observe the necessary crescent phase geometry to detect the glint effect. In general, the observability of glint depends on the distance to the system, the inclination-controlled phase angle range, and the inner working angle of the high contrast imaging system. While orbital inclination determines the range of phase angles that can be observed, the telescope's inner working angle, together with the planet's projected separation from its host star, limits which of those phases are accessible. Consequently, identifying the phase angle range over which glint can be observed constrains the orbital geometries and target systems for which ocean glint detection is feasible. Additionally, It is expected that spectral data quality (i.e., noise levels) should impact our ability to detect glint, in that noisier data will make glint detection more challenging. Thus, this study explores different retrieval scenarios on simulated HWO observations of an Earth-like, ocean-covered world at varying phase angles and signal-to-noise ratios (SNR).

While the work below emphasizes spectroscopy with non-polarimetric instruments, other studies have shown that polarimetry may provide a complementary avenue to surface ocean detection \citep{zugger2010light,trees2022ocean,roccetti2025planet}. Polarization can aid in glint detection by providing confirmation of polarization signatures associated with surface oceans \citep{mccullough2006modelspolarizedlightoceans,groot2020colors,stam2008spectropolarimetric,trees2022ocean}. At Brewster angle geometries for an air-water interface, (associated with phase angles of approximately $106^{\circ}$) the fraction of linearly polarized reflected light is expected to peak. This behavior could provide additional constraints on surface and atmospheric properties because the Brewster angle depends on the refractive indices of the two media.

Several complementary reflected-light techniques have been proposed to support the interpretation of exoplanet ocean detections. One complementary method is glint reddening. Because the optical path length increases at glancing geometries, Rayleigh scattering preferentially removes short-wavelength light during both the incoming and outgoing atmospheric paths. As a result, the reflected light glint spectrum becomes increasingly red-enhanced at larger phase angles \citep{ryan2022detecting}. Glint mapping is another complementary method that uses time-dependent variations in the glint signal as the planet rotates, allowing the distribution of oceans and continents to be constrained \citep{lustig2018detecting}. Early attempts to detect glint effects suggested that glint could be observed in Earth spatially-resolved and -unresolved whole-disk data \citep{sagan1993search,palle2003earthshine,robinson2014detection}. However more recent work indicates that aerosol scattering can mimic glint forward scattering in photometry \cite{Robinson_2026}. Further work suggests that spectroscopy may help to disentangle cloud scattering from glint \cite{cooper2025extreme}, but none of these works ran spectrally resolved inverse models, which is what our research sought to address. Thus, this work seeks to address the need for an exoplanet glint analysis tool by being the first to include glint in an inverse model to detect glint from HWO simulated results and using whole disk simulated spectra of Earth.

\section{Methods} \label{sec:style}

We updated an existing exoplanet retrieval tool, called \texttt{rfast}\citep{2023PSJ.....4...10R}, to include a treatment for ocean glint \citep{Cox:54}. In short, \texttt{rfast} pairs an efficient spectral forward model with a Markov chain Monte Carlo tool \citep[\texttt{emcee}; ][]{foreman2013emcee} to perform inverse analyses on real or synthetic planetary observations. Once the treatment was integrated into the model, we validated the tool. After validation, we ran a series of retrievals on simulated observations of a fully ocean-covered Earth-like exoplanet both with and without the glint treatment across a range of different phase angles and SNR. We then performed model selection on the ``with'' and ``without'' glint models to determine the phase angle and SNR at which the ocean glint signature was detectable. To help with visualizing example posteriors from our retrieval analyses, the Appendix demonstrates one corner plot from our inverse modeling grid. 

\subsection{rfast Updates} \label{subsec:tables}

The \cite{Cox:54} glint model uses wind speed to parameterize the statistical distribution of specularly-reflecting wave facet slopes, which determines the direction-dependent intensity reflection. The height of the waves producing the glint spot affect how diffuse the spot appears and depends on wind speed. We adopted a wind speed of 7\,m\,s$^{-1}$, informed by terrestrial global wind speed estimates \citep{archer2005evaluation} and previous exoplanet ocean-glint modeling studies \citep{trees2019blue}. We also ran forward models with higher and lower wind speeds of about 13\,m\,s$^{-1}$ and 3\,m\,s$^{-1}$ to confirm that brightening due to glint was similar for this range of wind speeds. Additionally, we compared forward models with high (75\%), medium (50\%) and low (25\%) cloud coverage to validate our updated tool against earlier results. To reproduce the high-fidelity phase-dependent Earth spectra of \citet{ryan2022detecting}, we adopted a cloud fraction of 75\%. This increased cloud fraction compensated for the absence of vertically optically thinner clouds in the current version of  \texttt{rfast} but that are treated in the high-fidelity model. For the retrieval experiments, however, we adopted a cloud fraction of 50\%, which is consistent with most previous exo-Earth retrieval studies \citep{2023PSJ.....4...10R, salvador2024influence,damiano2025effects}. 

The specular reflecting properties of a medium depends on its real and imaginary indexes of refraction, both of which depend on wavelength. However, across the wavelength range of interest (0.4--1.8\,$\upmu$m), the real index of refraction of water varies only by a small amount, and the imaginary index is generally quite small. We tested that a gray treatment for ocean glint was acceptable by calculating the bidirectional reflectance distribution function (BRDF) for the minimum and maximum indices of refraction in the wavelength range of interest and comparing the difference. The varied index of refraction across this wavelength range changed the output of the BRDF by a small enough amount that we concluded that the wavelength dependence did not need to be taken into account and a gray treatment could be used. For the model calculations, the  geometric means for both the real and imaginary indices of refraction in the range 0.4--1.8\,$\upmu$m were used. 


\begin{figure}
\centering
\includegraphics[width=0.492\textwidth]{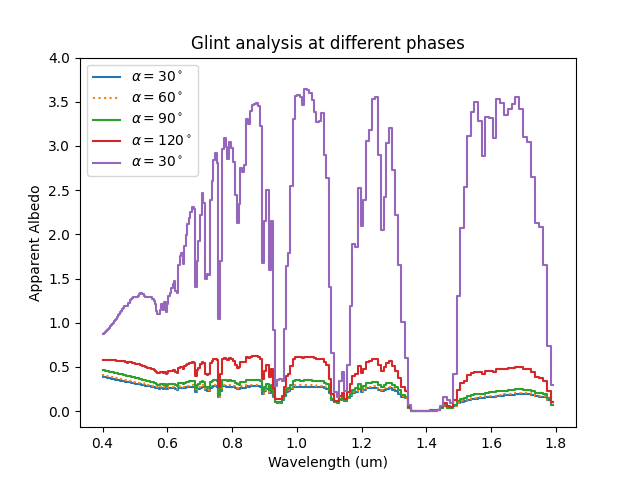}
\includegraphics[width=0.49\textwidth]{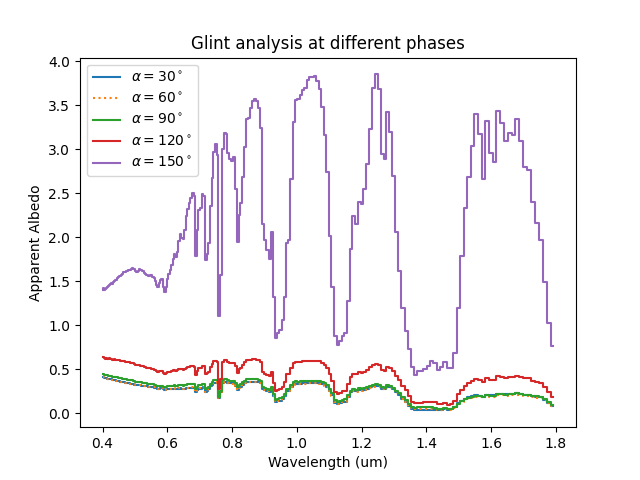}
\caption{Phase-dependent (colors) apparent albedo spectra from our updated \texttt{rfast} model for an ocean-covered Earth and assuming 75\% cloud coverage (left). The equivalent figure adapted from the high-fidelity Earth simulation data of \citet{ryan2022detecting} are shown for comparison (right). All spectra are at a resolving power ($\lambda/\Delta\lambda$) of 140. Deeper water bands in the \texttt{rfast} simulations are due to this model only including lower-atmospheric clouds, whereas the \citet{ryan2022detecting} simulations include high-altitude clouds that have lower contrast in gas absorption bands (as they obscure the deep atmosphere).}
\label{fig:earth_comparison}
\end{figure}

\subsection{Retrievals} \label{subsec:tables}

\begin{table}[H]
\centering
\begin{tabular}{@{} l c c c @{}}
\toprule
\textbf{Parameter} & \textbf{Symbol} & \textbf{Prior} & \textbf{Input Value} \\
\midrule

Surface pressure (Pa)
& $p_0$
& $\mathcal{LU}(0,\,8)$
& $\log(10^5)$ \\

Molecular nitrogen gas mixing ratio
& $f_{\rm N2}$
& $\mathcal{LU}(-10,\,0)$
& $\log(0.78)$ \\

Molecular oxygen gas mixing ratio
& $f_{\rm O2}$
& $\mathcal{LU}(-10,\,0)$
& $\log(0.21)$ \\

Water gas mixing ratio
& $f_{\rm H2O}$
& $\mathcal{LU}(-10,\,0)$
& $\log(0.003)$ \\

Carbon dioxide gas mixing ratio
& $f_{\rm CO2}$
& $\mathcal{LU}(-10,\,0)$
& $\log(4\times10^{-4})$ \\

Ozone gas mixing ratio
& $f_{\rm O3}$
& $\mathcal{LU}(-10,\,-2)$
& $\log(7\times10^{-7})$ \\

Methane gas mixing ratio
& $f_{\rm CH4}$
& $\mathcal{LU}(-10,\,0)$
& $\log(2\times10^{-6})$ \\

Planetary radius ($R_\oplus$)
& $R_{\rm p}$
& $\mathcal{LU}(-1,\,1)$
& $\log 1$ \\

Cloud coverage fraction
& $f_{\rm c}$
& $\mathcal{LU}(-3,\,0)$
& $\log 0.5$ \\

Column averaged temperature (K)
& $T_0$
& $\mathcal{U}(100,\,1000)$
& $294$ \\

Wind speed (m\,s$^{-1}$)
& $w$
& $\mathcal{U}(0.1,\,10)$
& $7$ \\

Cloud pressure extent (Pa)
& $\Delta p_{\rm c}$
& $\mathcal{LU}(0,\,8)$
& $\log(10^4)$ \\

Cloud top pressure (Pa)
& $p_{\rm t}$
& $\mathcal{LU}(0,\,8)$
& $\log(6\times10^4)$ \\

Cloud extinction optical depth
& $\tau_{\rm c}$
& $\mathcal{LU}(-3,\,3)$
& $\log(10)$ \\

\bottomrule
\end{tabular}
\end{table}

In order to constrain the necessary observational parameters for ocean glint detection, we created a grid of simulated HWO spectra of an Earth-like world over different phase angles and SNR. For simplicity and clarity we assumed that the surface is fully ocean-covered, which implies that we are over-estimating rotationally-averaged glint contributions by 10--30\% as compared to the true Earth (the true value of the over-estimation would depend on viewing geometry). Retrievals were then performed on these mock observations for cases where the inverse model both did and did not include ocean glint. Instead of a glint treatment, the glint-free models adopted a Lambertian surface reflection treatment where the gray surface flux albedo was a fitted parameter. We focused on a range of SNRs from 5-25 in increments of 5 and a range of phase angles from $90^{\circ}$ to $150^{\circ}$ in increments of 10$^{\circ}$. In effect, our retrievals aim to address whether high-phase spectral observations of Earth can be sufficiently fit with a cloud contribution and a Lambert surface or if a specular-like surface contribution is required. In order to intercompare the two inverse models (glinting and non-glinting), we used the Bayesian Information Criterion (BIC), a common statistical metric for model comparison in astronomy \citep{tribbett2020titan,cornish2007tests}. The BIC is defined as,
\begin{equation}
{\rm BIC} = {-2 \ln P (y|\theta)}_{\rm max} + {\nu \ln N} \ ,
\end{equation}
where ${P (y|\theta)}_{max}$ is the maximized log likelihood, $\nu$ is the number of free parameters in the model, and $N$ is the number of observational data points. This expression can be related to the reduced chi-squared ($\chi^{2}_{\nu}$) using,
\begin{equation}
{\rm BIC} = (N-\nu) \chi^{2}_{\nu} + {\nu \ln N} \ .
\end{equation}
Generally, models with a smaller BIC value are preferred, so we adopted the $\Delta$BIC as a means of statistical comparison between models. Here, a $\Delta$BIC between 0 and -2 means the difference between the two models is not worth more than a bare mention, a $\Delta$BIC between -2 and -6 corresponds to positive evidence in favor of the model with the smaller BIC value, values between -6 and -10 indicate strong evidence, and values less than -10 indicate very strong evidence in favor of the model with the smaller BIC value \citep{kass1995bayes}.(Note that some explanations and applications of the BIC flip the convention so that larger positive $\Delta$BIC values correspond to preferred models.) For our calculations of the $\Delta$BIC, we had 14 free parameters ($\nu$) and 129 wavelength data points ($N$) and $\chi^2$ differed for the with-glint and without-glint cases and was calculated within the \texttt{rfast} code. 

\section{Results} \label{sec:floats}

\begin{figure} 
    \centering
    \includegraphics[width=0.5\linewidth]{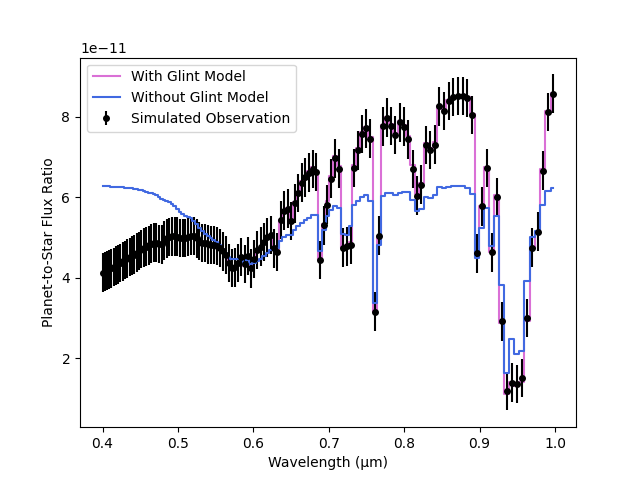}
    \caption{Simulated HWO data and error bars (black) alongside a best-fit glinting model (pink) and non-glinting model (blue) for phase angle of 135$^{\circ}$ and SNR of 10. Retrievals that include glint are decisively preferred to the retrievals that do not include glint.}
    \label{fig:demo}
\end{figure}

Our characterization analysis investigated both individual retrieval comparisons as well as the general behaviors of glint detectability with phase angle and SNR. Figure~\ref{fig:demo} demonstrates retrieval results from a single SNR-level and phase angle pair in our larger grid. A simulated HWO observation (black points) at a phase angle of 135$^{\circ}$ and a characteristic SNR of 10 can be well-fit by a model that includes ocean glint (pink). However, even the best-fit result from a glint-free retrieval (blue) cannot match the glint-reddened shape of the observation. Thus, this particular SNR and phase angle pair yields a very strong detection of an ocean glint contribution.

\begin{figure} 
    \centering
    \includegraphics[width=0.5\linewidth]{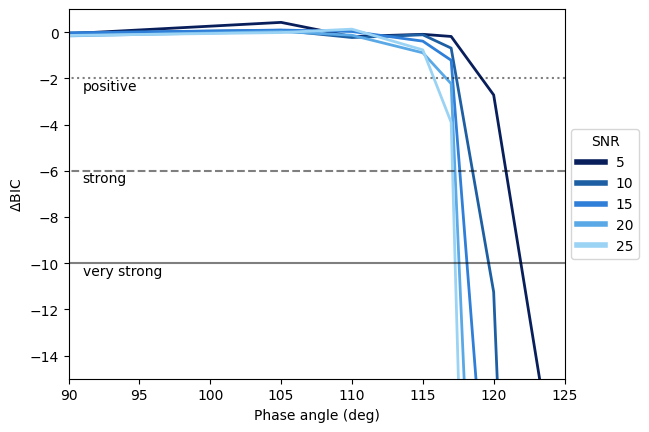}
    \caption{Model selection results (quantified as $\Delta {\rm BIC}$) as a function of observed phase angle and for different SNRs (colors). Regardless of data quality, the glinting model is preferred to the non-glinting model for phase angles of about $120^{\circ}$ and above.}
    \label{fig:dbic}
\end{figure}

Figure~\ref{fig:dbic} shows our model selection results using the $\Delta$BIC as a statistical comparison for our grid of phase angles and SNRs. Most generally, the glinting model is preferred to the non-glinting model at phase angles of about $120^{\circ}$ and above, regardless of the noise level. Larger SNRs require access to a slightly smaller phase angle to detect the glint contribution, and vice versa.


\begin{figure} 
    \centering
    \includegraphics[width=0.5\linewidth]{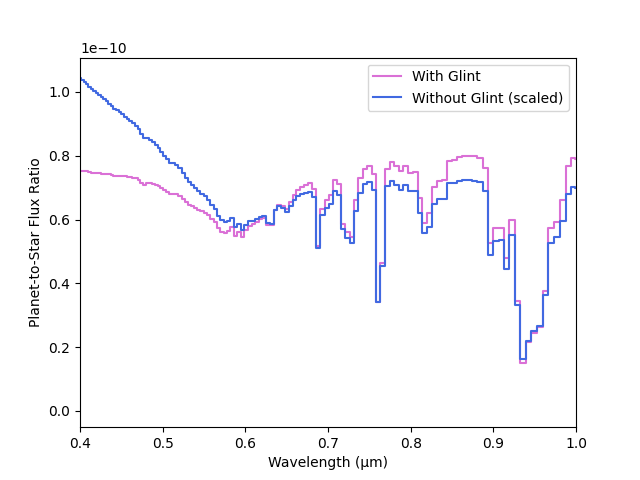}
    \caption{Comparison of glinting (pink) and non-glinting (blue) forward models at a phase angle of $120^{\circ}$. The latter is scaled by 1.5x to best enable the by-eye comparison of the spectral shape.}
    \label{fig:explain}
\end{figure}

Figure~\ref{fig:explain} compares glinting and glint-free forward models at a phase angle of $120^{\circ}$, which helps to explain the significance of this phase. Specifically, comparing these two spectra shows that this phase angle is where (in our models) the spectral shape changes from red to blue when glint is removed. Furthermore, the characteristic difference between the ``with glint'' and ``without glint'' spectra is 10--20\%, in-line with the smallest SNRs we explored. We note that the glint-free model was scaled to better enable a comparison of the spectral shapes, where\,---\,in a retrieval\,---\,such a scaling can be obtained by biasing the radius or cloud fraction. At shorter, bluer wavelengths, the unscaled ``without glint'' case has a comparable planet-to-star flux ratio to the ``with glint'' case as the surface sensitivity is reduced at these wavelengths by Rayleigh scattering.

    
\section{Discussion} \label{sec:floats}

\subsection{Transition from Non-Detection to Detection}

The ubiquitous glint detection at a phase angle of $120^{\circ}$ in our model selection results is associated with a very steep transition from non-detection to detection. As can be seen in Figure~\ref{fig:dbic}, there is a very strong preference for the glinting model over the non-glinting model that arises across all the SNRs at the $120^{\circ}$ phase angle. Higher SNRs lead to a glinting model preference arising at smaller phase angles than for low SNRs. The spectral SNR did not strongly impact the detectability at phase angles at (or above) $120^{\circ}$, although the ${\rm SNR}=5$ retrievals show a weaker glinting model preference (i.e, smaller $\Delta$BIC) than the other larger SNR cases. This suggests that retrievals run at yet-smaller SNRs may yield glint non-detections at a phase angle of $120^{\circ}$. The physical process that leads to the sharp increase in detectability at $120^{\circ}$ is the pronounced increase in glint reddening between phase angles of $110^{\circ}$--$120^{\circ}$. Figure~\ref{fig:explain} shows that the shape of the spectrum (i.e., whether the spectrum is blue versus red) is quite sensitive to whether a glint component is included at (and above) $120^{\circ}$.

\subsection{HWO Observability}

For an Earth-Sun twin system located at a distance of 10\,pc and viewed at an orbital inclination of i=$90^{\circ}$, the maximum projected star--planet separation is 100 mas. The minimum phase angle for glint detection identified in this study, $120^{\circ}$, corresponds to a projected separation of 86.6 mas (100$\sin120^{\circ}$). This result is encouraging for HWO observations. Assuming a telescope diameter of 6 m, an inner working angle of even 4$\lambda/D$ at 0.6\,$\upmu$m would permit observations at this geometry and distance (assuming a favorable inclination). At a wavelength of 1\,$\upmu$m (the longest wavelength investigated in this work), an inner working angle of 2.5$\lambda/D$ would be required to reach the requisite planet-star separation for this target at 10\,pc.

The relationship between phase angle, orbital position, and orbital inclination (relative to the plane of the sky) enables us to determine the range of inclinations where the planet can reach a given extreme in phase, with, 
\begin{equation}
    \cos(\alpha)=\sin(\theta+\omega_p)\sin(i) \ ,
\end{equation}
where $\theta$ is the true anomaly, $\omega_p$ is the argument of periastron, and $i$ is the orbital inclination \citep{ryan2022detecting}. As the first sine term on the right-hand-side is bounded to be between $-1$ to 1, the extremes of the right-hand-side are $\pm \sin(i)$. With $\cos(120^{\circ})=-1/2$, we find that inclinations between 30$^{\circ}$--150$^{\circ}$ allow for the planet to reach the requisite phase angle for confident glint detections. The phase angle lower limit for which ocean glint is detectable has previously been estimated to be $130^{\circ}$ \citep{vaughan2023chasing}, but this work supports that glint may be detectable over a larger range (down to $120^{\circ}$) of phase angles.

Using a preliminary HWO list of potential candidates is comprised of 164 stars within 25 pc from Earth \citep{2024arXiv240212414M} that could potentially host rocky planets suitable for characterization, \citet{vaughan2023chasing} averaged over 1,000 simulations with randomized inclinations to calculate the expected number of most extreme phase angles accessible for different IWA. They also calculated the number of Habitable Zone Earth-like planets that could be imaged at these phase angles, assuming an occurrence rate of $\eta_\oplus$ = 0.24. The results indicate that for an IWA of 21 mas, a maximum accessible phase angle of $120^{\circ}$ corresponds to approximately 145 accessible systems and 35 planets. Alternatively, an IWA of 85 mas and a maximum accessible phase angle of $120^{\circ}$ corresponds to approximately 50 systems and 10 planets. 
While our current work does not explore the exposure times that would be required to achieve the quality of spectral observations we assume in our retrievals, we note that \citet{ryan2022detecting} found exposure times for glint detections for Earth-twin targets at 8\,pc and with inclinations larger than 30$^{\circ}$ of roughly 200\,hr for the HabEx mission concept \citep{gaudi2020habitableexoplanetobservatoryhabex} and 3--40\,hr for the LUVOIR-A (15\,m) concept \citep{theluvoirteam2019luvoirmissionconceptstudy}.


\subsection{Future Work}

Several simplifying assumptions were adopted in this study, including a fully ocean-covered Earth twin, fixed atmospheric composition, prescribed cloud distributions (50\%), and idealized surface properties. Real exoplanets may exhibit heterogeneous surfaces, variable cloud cover, hazes, and seasonal variability, all of which could affect glint detectability. Future work could modify the assumptions adopted in the study to better constrain how these variables effect glint detection. In general, though, we expect that the glint detectability will decrease as increasingly more the of surface is covered by land or is obscured by clouds or hazes, where non-detections would then lead to false-negatives.


As for false positives, hazes have been shown to lead to glint-like forward scattering from Titan \citep{2017NatAs...1E.114G,cooper2025extreme}. Additionally, recent inverse modeling work on photometric observations of Earth's phase curve showed that vertically optically-thin aerosols (e.g., some cirrus clouds) can mimic Earth's glint signature at visual wavelengths \citep{Robinson_2026}. It is possible that these aerosol forward scattering false positive scenarios could be differentiated from glint detection through spectral comparison. More specifically, \citet{cooper2025extreme} showed that the extent of haze forward scattering is enhanced in/near gas absorption bands, as multiply-scattered radiation at these wavelengths tends to be absorbed, thus leaving mainly single-scattered (i.e., strongly forward scattering) light. This is opposite of the prediction for glint, where such ocean scattering will be reduced in/near gas absorption bands due to limited surface sensitivity at these wavelengths \citep{robinson2010detecting,zugger2010light}. An avenue for future work would be to update \texttt{rfast} (or any equivalent inverse model) to allow for multiple cloud layers and thicknesses. Performing retrievals and model selection in a similar fashion to the work above would demonstrate if forward scattering from clouds/hazes can mimic even spectral glint signatures. Finally, performing retrievals on actual crescent-phase spectral observations of Earth (which, at the time of this writing, are exceedingly rare) would be a key validation of any glint inference methods.


\section{Conclusion} \label{sec:floats}

The spectral detection of ocean glint signatures could be an essential method for constraining the habitability of distant exoplanets for the Habitable Worlds Observatory and other future direct imaging missions. Our main results are as follows:

\begin{itemize}
    \item We updated a retrieval model to include an ocean glint treatment and validated this tool against a high-fidelity ``Earth as an exoplanet'' model. The updated and validated retrieval tool is now well-suited to glint detectability studies for HWO.

    \item We used model selection techniques for retrievals that did, and did not, include a glint component. These retrievals were performed on a grid of simulated, HWO-like observations that spanned 90$^{\circ}$--150$^{\circ}$ in phase angle (i.e., crescent phases) and spectral SNRs of 5--25. 

    \item Model selection results showed that the glint model is preferred over the non-glint model at phase angles of $120^{\circ}$ and greater, regardless of the spectral SNR. Although higher SNRs produce increasingly decisive statistical preferences for the glinting model at $120^{\circ}$, they do not substantially reduce the minimum phase angle at which glint can be detected.
 
\end{itemize}

\begin{acknowledgments}
EC is grateful for funding support from the Arizona Astrobiology Center. Additionally, the material contained in this document is based upon work supported by a National Aeronautics and Space Administration (NASA) cooperative agreement 80NSSC20M0041. Any opinions, findings, conclusions or recommendations expressed in this material are those of the author and do not necessarily reflect the views of NASA. This work was supported through a NASA grant awarded to the Arizona/NASA Space Grant Consortium. EC and TDR were both supported by NASA's Exoplanets Research Program (No. 80NSSC25K7149). TDR also appreciates funding support from NASA's Nexus for Exoplanet System Science Virtual Planetary Laboratory (No.~80NSSC23K1398).

\textit{Software:} \texttt{emcee} Markov Chain Monte Carlo sampler\citep{foreman2013emcee}, \texttt{corner} \citep{foreman2016corner}, \texttt{rfast} \citep{2023PSJ.....4...10R}

\end{acknowledgments}

%



\appendix

\begin{figure}[H] 
        \centering
        \includegraphics[width=0.7\linewidth]{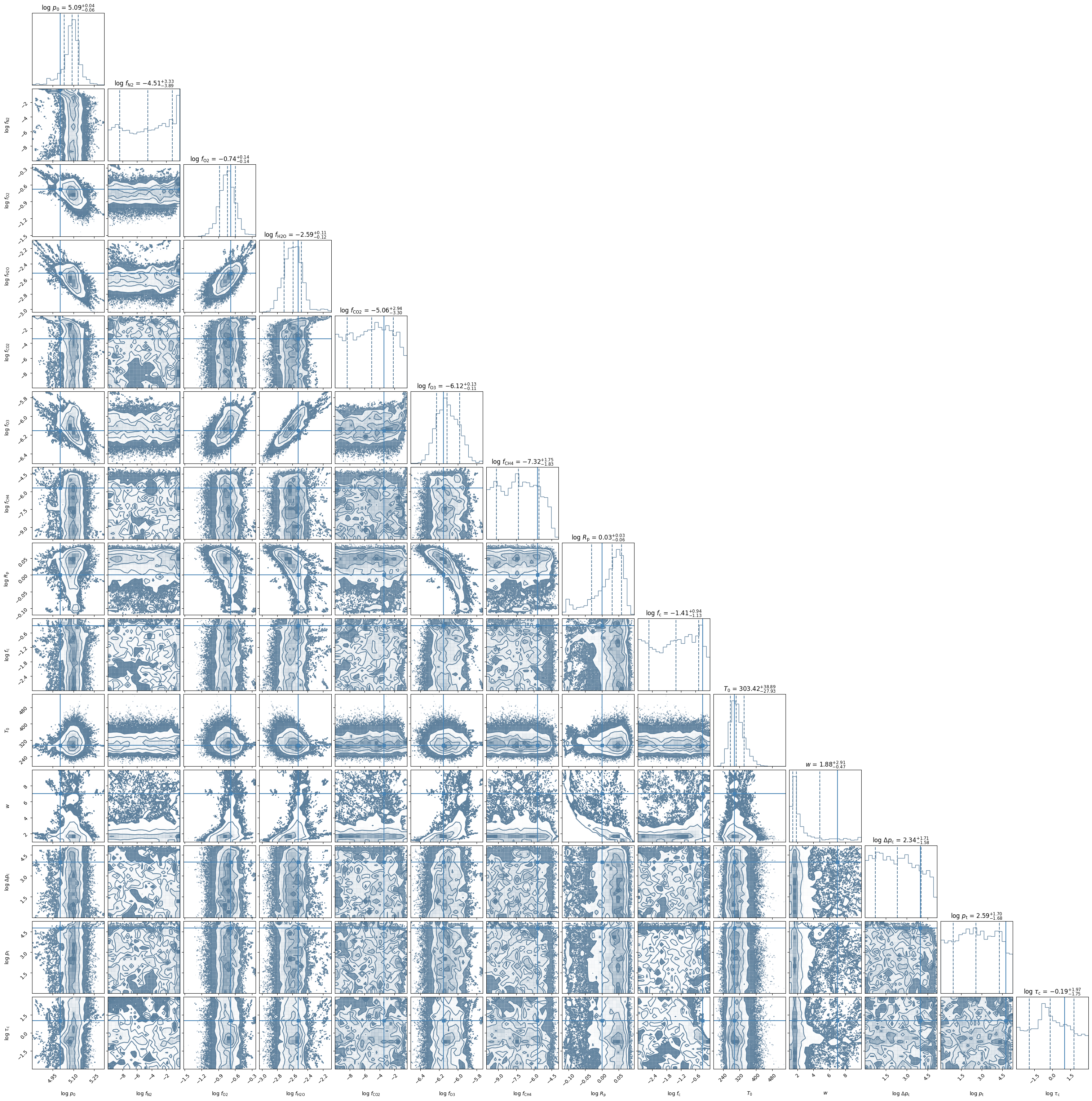}
        \caption{Example corner plot from SNR 10, phase angle $130^{\circ}$ retrieval, including glint. The vertical line indicates the characteristic ``truth'' value for Earth. Confidence intervals at the 16/50/84-the percentiles are indicated and provided at the top of the one-dimensional marginal distributions along the diagonal.}
        \label{fig:corner}
    \end{figure}

Figure~\ref{fig:corner} demonstrates an example corner plot \citep{foreman2016corner} from the SNR 10 and $135^{\circ}$ phase retrieval, including inferred glint-related quantities. Running along the diagonal the fitted parameters are surface pressure ($p_0$) gas mixing ratios for molecular nitrogen, molecular oxygen, water, carbon dioxide, ozone and methane (fN2, fO2, fH2O, fCO2,fO3, and fCH4, respectively; assumed to have constant vertical profiles), planetary radius ($R_p$), cloud coverage fraction on the planetary disk ($f_c$), column averaged temperature($T_0$), wind speed ($w$), cloud pressure extent ($\Delta p_c$ ), cloud top pressure in Pa ($p_t$), cloud extinction optical depth ($\tau_c$). 

\end{document}